# *High-order Large Eddy Simulations of Confined Rotor-Stator Flows*

S. Viazzo*, S. Poncet, E. Serre, A. Randriamampianina & P. Bontoux

*Laboratoire M2P2, UMR 6181 CNRS / Aix-Marseille Université,*

*IMT la Jetée, 38 rue F. Joliot-Curie, 13451 Marseille, cédex 20, France*

* Tel. +33 (0) 4.91.11.85.55, Fax +33 (0) 4.91.11.85.02,
stephane.viazzo@L3m.univ-mrs.fr

**Abstract**

In many engineering and industrial applications, the investigation of rotating turbulent flow is of great interest. In rotor-stator cavities, the centrifugal and Coriolis forces have a strong influence on the turbulence by producing a secondary flow in the meridian plane composed of two thin boundary layers along the disks separated by a non-viscous geostrophic core. Most numerical simulations have been performed using RANS and URANS modelling, and very few investigations have been performed using LES. This paper reports on quantitative comparisons of two high-order LES methods to predict a turbulent rotor-stator flow at the rotational Reynolds number Re($=\Omega b^2/\nu$)=$4\times10^5$. The classical dynamic Smagorinsky model for the subgrid-scale stress (Germano et al. 1991) is compared to a spectral vanishing viscosity technique (Séverac & Serre 2007). Numerical results include both instantaneous data and post-processed statistics. The results show that both LES methods are able to accurately describe the unsteady flow structures and to satisfactorily predict mean velocities as well as Reynolds stress tensor components. A slight advantage is given to the spectral SVV approach in terms of accuracy and CPU cost. The strong improvements obtained in the present results with respect to RANS results confirm that LES is the appropriate level of modelling for flows in which fully turbulent and transition regimes are involved.

## 1 Introduction

The simulation of turbulent rotating cavity flows is a major issue in computational fluid dynamics and engineering applications such as computer disk drives, automotive disk brakes, and disks to support turbomachinery blades. Besides its primary concern to industrial applications (Owen & Rogers 1989), the rotor-stator problem has also proved a fruitful means of studying the effects of mean flow three-dimensionality on turbulence and its structure (Littell & Eaton 1994, Lygren & Andersson 2004, Séverac et al. 2007), because it is among the simplest flows where the boundary layers are three-dimensional from their inception. There is no effect of inflow condition (turbulence level) and the flow can also involve multiple regimes such as laminar, transitional, fully turbulent and relaminarization. A renewed interest for these flows was born from the appearance of precessing large scale vortical structures embedded in the turbulent regime, first observed experimentally by Czarny et al. (2002) and confirmed numerically by the URANS computations of Craft et al. (2008).

RANS modelling of rotating turbulent flows has been investigated by the group of B.E. Launder in Manchester for several decades (see a review in Launder et al.

2010). Conclusions of their investigations are that eddy-viscosity models clearly fail, and provide erroneous rotor layer predictions and rotation rates in the central core. This is particularly due to a delay in the transition to turbulence along the rotor (Iacovides & Theofanoupolos 1991). Due to the skewing of the boundary layers, the Reynolds stresses are not aligned with the mean flow vector. This invalidates the assumptions of eddy-viscosity models. Second moment closures provide a more appropriate level of modelling (Launder & Tselepidakis 1994; Poncet et al. 2005b), but even if they provide a correct distribution of laminar and turbulent regions, the Reynolds stress behaviour is not fully satisfactory, particularly near the rotating disk.

Consequently, LES seems to be the appropriate level of modelling. Wu and Squires (2000) performed the first LES of the three-dimensional turbulent boundary layer over a free rotating disk in an otherwise quiescent incompressible fluid at $Re = 6.5 \times 10^5$. Periodic boundary conditions were used both in the radial and tangential directions. They concluded that when the grid is fine enough to accurately resolve large-scale motions, SGS models and further grid refinement have no significant effect on their LES predictions. Lygren and Andersson (2004) and Andersson and Lygren (2006) performed LES of the axisymmetric and statistically steady turbulent flow in an angular section of an unshrouded rotor-stator cavity for Reynolds numbers ranging from $Re = 4 \times 10^5$ to $Re = 1.6 \times 10^6$. Their study showed that the mixed dynamic subgrid scale model of Vreman et al. (1994) provided better overall results compared to the dynamic subgrid scale model of Lilly (1992). Finally, Séverac and Serre (2007) proposed a LES approach based on a spectral vanishing viscosity (SVV) technique, which was shown to provide very satisfactory results with respect to experimental measurements for an enclosed rotor-stator cavity including confinement effects and for Reynolds numbers up to $Re = 10^6$ (Séverac et al. 2007). Clearly, traditional LES approaches using a constant subgrid-scale eddy-viscosity are not able to predict the transition to turbulence on the rotor layer. In this work, the dynamic eddy-viscosity SGS model of Lilly (1992) is used with a fourth-order compact finite-difference scheme, denoted LES-FD in the following. LES-FD results are then compared with spectral vanishing viscosity results formerly published in Séverac et al. (2007) for the same configuration. This alternative LES formulation (LES-SVV) used a spectral approximation based on a modification of the Navier–Stokes equations. The LES-SVV model only dissipates the short length scales, a feature which is reminiscent of LES models, and keeps the spectral convergence of the error (Séverac & Serre 2007).

The current work evaluates the above two LES methods in a very complex configuration including both laminar and turbulent flow regions with persisting unsteady structures and involving confinement. Moreover, flow curvature and rotation effects provide a strongly inhomogeneous and anisotropic turbulence within the boundary layers along the disks and the endwalls. The use of high-order numerical approximations guarantees that the dissipative truncation error (which is nearly zero) cannot act as an additional subgrid model.

This article is organized as follows. First geometrical and numerical models are briefly described in Sections 2 and 3, respectively. Comparisons between the LES calculations and previous velocity measurements (Séverac et al. 2007) are performed for a given Reynolds number $Re=4 \times 10^5$ in Section 4 for the mean and turbulent fields. Finally some conclusions and closing remarks are provided in Section 5.

## 2 Geometrical Modelling

The cavity considered in the current study is illustrated in Figure 1. It is composed by two parallel disks of radius $b=140$mm. The lower disk rotates at a uniform angular velocity $\Omega$ (rotor), while the upper disk is at rest (stator). The disks are delimited by an inner cylinder (the hub) of radius $a=40$ mm attached to the rotor by an outer stationary casing (the shroud) attached to the stator. The interdisk spacing, denoted $h$, is fixed to 20 mm.

The mean flow is governed by three global control parameters: the aspect ratio of the cavity $G$, the curvature parameter $R_m$ and the rotational Reynolds number Re based on the outer radius $b$ of the rotating disk defined as follows:

$$G = \frac{b-a}{h} = 5, \quad R_m = \frac{b+a}{b-a} = 1.8, \quad Re = \frac{\Omega b^2}{\nu} = 4\times 10^5 \quad (1)$$

where $\nu$ is the fluid kinematic viscosity. The values of the geometrical parameters are chosen to be relevant with industrial devices, such as the stages of a turbopump, and to satisfy technical constraints of experimental devices as well as computational effort to reach statistically converged states. In the experimental setup, the clearance which exists between the rotor and the shroud, is small and equal to 0.85 mm. In the following, the stator is located at $z^*=z/h=1$ and the rotor at $z^*=0$. We also define the dimensionless radial location as $r^*=(r-a)/(b-a)$ varying between 0 (at the hub) and 1 (at the shroud).

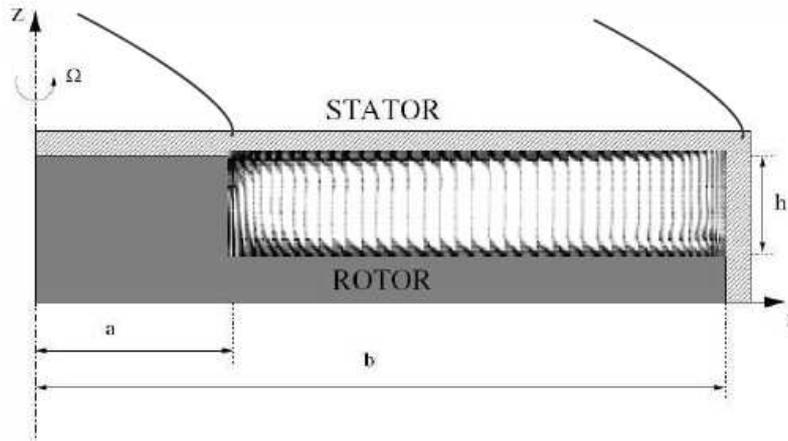

**Fig. 1** The geometry of the rotor-stator cavity and relevant notations. The averaged velocity vector-field (LES-SVV) shows the secondary flow in the meridian plane at Re=$4\times 10^5$.

## 3 Numerical set up

The incompressible fluid motion is governed by the three-dimensional Navier-Stokes equations written in primitive variables for cylindrical coordinates ($r$, $\theta$, $z$). No-slip boundary conditions are applied at all walls so that all the near-wall regions are explicitly computed. The tangential velocity is fixed to zero on the stator and on the shroud and to the local disk velocity $\Omega r$ on the rotor and the hub. The physical discontinuities between rotating and stationary walls are regularized in both LES codes. The tangential velocity profile is smoothed using an exponential function.

Conservation equations are solved using a Fourier approximation in the homogeneous tangential direction. In both non homogeneous radial and vertical directions the solutions are approximated using either a fourth-order compact finite-difference scheme (LES-FD; see in Abide & Viazzo 2005) or a collocation-Chebyshev approximation (LES-SVV; see in Séverac & Serre 2007).

In both numerical codes, the time advancement is second-order accurate and is based on the explicit Adams–Bashforth time-stepping for the convective terms and an implicit backward-Euler scheme for the viscous terms. The velocity-pressure coupling is solved using a two-step fractional scheme (predictor–corrector) reducing the problem at each time step to a set of two-dimensional Helmholtz equations (Raspo et al. 2002).

In the LES-FD, mass and momentum conservations are enforced for the large-scale resolved variables, by filtering the Navier-Stokes equations. The unresolved small scale turbulence is modelled by a dynamic eddy-viscosity SGS model proposed by Lilly (1992) in which the Smagorinsky coefficient $C_s$ is evaluated as a part of the solution at each time step using a test filter. As is common practice in the literature, the test filter width was twice the grid-filter width. The test filter used sequentially in each non-homogeneous direction is a symmetric discrete filter based on the trapezoidal rule as proposed in Sagaut (2005), while in the homogeneous direction a cut-off filter was applied with the Fourier approximation. Furthermore, when using dynamic Smagorinsky model, negative values of $\nu_T$ are set to zero to avoid numerical instabilities. As a consequence, no transfer of energy from the unresolved scales to the resolved ones (backscatter) is taken into account. This is consistent with the LES-SVV model. In order to improve time-stability, the eddy viscosity is split into an averaged value in the azimuthal direction treated semi-implicitly by the way of internal iterations and a fluctuation part treated fully explicitly. In Practice, five iterations are required to obtain a convergence criterion of $10^{-6}$. Finally, as expected from the dynamic procedure, the eddy viscosity acts more significantly at large radii where the turbulence intensity is higher and vanishes in the laminar regions (Figure 2).

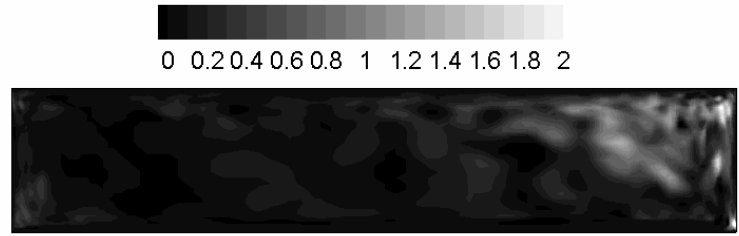

**Fig. 2** Isocontours of the ratio of the averaged eddy viscosity to the molecular viscosity, $\nu_t/\nu$, in the (r, z) plane calculated using the LES-FD method.

In the LES-SVV, an appropriate viscosity kernel operator is incorporated in the conservation equations. This operator is only active for high wave numbers. It does not affect the large scales of the flow and stabilizes the solution by increasing the dissipation, particularly near the cut-off frequency, without sacrificing the formal accuracy, i.e., exponential convergence (Séverac & Serre 2007). A new diffusion operator $\Delta_{SVV}$ can be simply implemented by combining the classical diffusion and the new SVV terms to obtain:

$$\nu \Delta_{SVV} = \nu \Delta + \nabla.(\varepsilon_N Q_N \nabla) = \nu \nabla . S_N \nabla \qquad (2)$$

$$S_N = diag\{S_N^i\} \; S_N^i = 1 + \frac{\varepsilon_N^i}{\nu} Q_N^i \qquad (3)$$

where $\varepsilon_{N_i}^i$ is the maximum of viscosity and $Q_N^i$ is a 1D viscosity kernel operator acting in direction i =1, 2, 3 (corresponding to *r*, θ or *z*), and defined in the spectral space by an exponential function:

$$\hat{Q}_N(\omega_n) = 0 \text{ for } 0 \leq \omega_n \leq \omega_T \tag{4}$$

$$\hat{Q}_N(\omega_n) = \varepsilon_N e^{-[(\omega_N - \omega_n)/(\omega_T - \omega_n)]^2} \text{ for } \omega_T \leq \omega_n \leq \omega_N \tag{5}$$

where $\omega_T^i$ is the threshold after which the viscosity is applied and $\omega_N^i$ is the highest frequency calculated in the direction *i*. Note that the SVV operator affects at most the two-third of the spectrum on the highest frequencies ($\omega_T = 0$). Consequently, DNS results are easily recovered for laminar flows. Moreover, the SVV term is not scaled by the Reynolds number. Therefore, for a fixed grid and given SVV parameters, the SVV term may become larger relative to the classical diffusion term when increasing the Reynolds number.

Because our SVV operator is fully linear, no additional computational cost is needed. According to Séverac & Serre (2007), the control parameters used in the present LES-SVV are summed up in table 1 and behave as $\varepsilon_{N_i}^i = O(1/N_i)$ and $\omega_T^i = O(N_i^{1/2})$.

| direction | $\omega_T$ | $\varepsilon_N$ | Grid |
|---|---|---|---|
| *r* | $0.8 N^{1/2}$ | $1/(2N)$ | *N*=121 |
| θ | $N^{1/2}$ | $1/(2N)$ | *N*=181 |
| *z* | $N^{1/2}$ | $1/(2N)$ | *N*=65 |

**Table 1** LES-SVV control parameters and grid. $\varepsilon_N$ is the amplitude and $\omega_T$ the frequency threshold.

The LES-SVV solver has been fully validated numerically and physically in Séverac & Serre (2007) and in Séverac et al. (2007) in the same geometrical configuration as the current study. The LES-FD cartesian solver has been fully validated in Beaubert & Viazzo (2003) and in Abide & Viazzo (2005). The numerical validation of the cylindrical solver is done here through the satisfactory agreement outlined in the cross-comparisons presented in the following section.

The spatial resolution and accuracy have been checked *a priori* and *a posteriori*. The computational grid 121×65×180 in the *r*, *z*, and θ directions has been sized for the two methods taking into account both the computational resource and the *a priori* estimation obtained from our former work in the same configuration for Reynolds numbers ranging from $10^5$ to $10^6$. LES-SVV calculations have been performed in the whole cavity using a time step $\delta t = 5.10^{-5}$ $\Omega^{-1}$ whereas LES-FD has been only performed in a half-cavity [0, π] and a time step $\delta t = 3.10^{-4}$ $\Omega^{-1}$ in order to save CPU-time when using the costly dynamics procedure. Former calculations of Séverac et al. (2007) at Re=$10^6$ have shown that forcing π-periodic solutions could remain acceptable, because the two-point correlations in this direction are nearly zero over this distance.

An *a priori* verification of the resolution has been performed by calculating an estimation of the ratio of the resolved to modelled scales, i.e. Δmax/ η where Δmax is the maximum cut-off wavelength given by the grid and η the Kolmogorov length scale estimated from the homogeneous isotropic turbulence theory using measurements of the size and velocity of the primary rolls given by linear stability analysis in Serre et al. (2004). In LES-SVV and LES-FD calculations, the ratio $\Delta$max /$\eta$ is around 2.5 and 3.3 respectively. This permits to resolve near-wall structures. The numerical resolution has been checked *a posteriori* by calculating the friction velocity in both boundary layers and in the core. Both LES-SVV and LES-FD grids correspond to an axial wall-coordinate $z^+$ around unity slightly varying with $r$ as the total friction velocity increases towards the periphery of the cavity whereas $r^+$ and $\theta^+$ are around 10 at mid-cavity.

After the time-dependent flow is settled at a statistically steady state, turbulence statistics are averaged both in the homogeneous tangential direction and in time during around 10 global time units (in term of $\Omega^{-1}$). The state is considered statistically steady when the fluctuations of the averaged values in time are less than 1%.

# 4 Results

LES results have been compared to velocity measurements using a two component laser Doppler anemometer from above the stator (see details in Séverac et al. 2007). About 5000 validated data are necessary to obtain the statistical convergence of the measurements. The mean and turbulent quantities are given with an accuracy of 2% and 5% respectively. Comparisons with the sophisticated Reynolds Stress Modelling (RSM) of Elena and Schiestel (1996) are also provided as a reference to RANS modelling. This model has been sensitized to the implicit effects of rotation on turbulence and applied to a wide range of flow conditions in enclosed or opened rotor-stator cavities (Poncet et al. 2005b).

Both LES-FD and LES-SVV provide the same axisymmetric base flow. It is characterized by a secondary flow in the meridian plane induced by the centrifugal force as illustrated by the averaged velocity vector field in Figure 1. As a consequence of the Coriolis force (Taylor-Proudman theorem), the flow is composed of two thin boundary layers along the disks separated by a geostrophic core at zero axial pressure gradient. The fluid is pumped centrifugally outwards along the rotor and is deflected in the axial direction after impingement on the external cylinder. After a second impingement on the stator, it flows radially inwards along the stator, due to conservation of mass, before turning along the hub and being centrifuged again by the rotating disk. Note that due to a smaller radial velocity, the Bödewadt layer along the stator is almost twice as thick as the Ekman layer on the rotor. The thicknesses of both boundary layers are scaled by the characteristics viscous length $\delta=(\nu/\Omega)^{1/2}$.

The radial confinement by both cylinders makes the solutions inconsistent with self-similarity solutions, although they can be close far from the endwalls. In particular, both boundary layers are not parallel. Thus, comparisons between LES results and available linear stability results can be local only.

In consequence, quantitative comparisons between LES-SVV and LES-FD have been made at three radial locations of the cavity $r^*=0.3$, $r^*=0.5$ (mid-radius) and $r^*=0.7$, that correspond to a local Reynolds number $Re_r =(r/\delta)$ varying in the range $10^5 \leq Re_r \leq 2.46\times 10^5$.

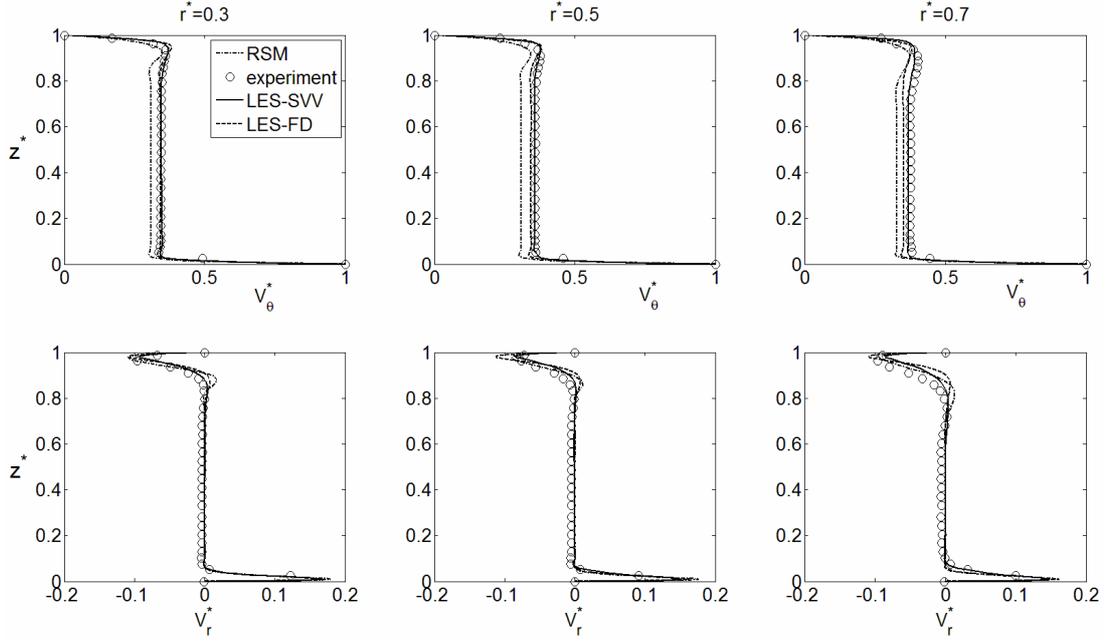

**Fig. 3** Mean profiles of tangential and radial components of velocity normalized by the local speed of the disk at three radial locations. Comparisons between the LES-SVV (full lines), the LES-FD (dashed lines), the velocity measurements (circles) and the RSM model (dash-dotted lines).

Figure 3 shows the axial profiles of the mean radial $V_r^*=V_r/(\Omega r)$ and tangential $V_\theta^*=V_\theta/(\Omega r)$ velocity components normalized by the rotor velocity at these radial locations. The mean axial velocity component is not shown here because it is nearly zero in this range of radii far from the endwalls.

The agreement between experimental measurements and both LES predictions is satisfactory. Nevertheless, LES-SVV provides better overall results than the LES-FD and the RSM. The boundary layer thicknesses are globally well predicted. Both LES methods slightly underestimate thicknesses, especially at large radii for LES-FD. The velocity maxima within the boundary layers are well predicted by both LES over the stator but strongly overestimated over the rotor. The core swirl ratio or entrainment coefficient $K=(V_\theta/(\Omega r))_{core}$ is crucial for engineering applications because it is directly linked to the radial pressure gradient in the cavity and consequently to the axial thrusts applied on the rotor (Poncet et al. 2005a). The LES-FD slightly underestimates $K$, predicting $K=0.345$ at $r^*=0.5$ with respect to $K=0.36$ given by LES-SVV as well as experimental measurements. This underestimation is more pronounced by the RSM which predicts $K=0.315$ close to the value for laminar similarity solutions ($K=0.313$) (Pearson 1965).

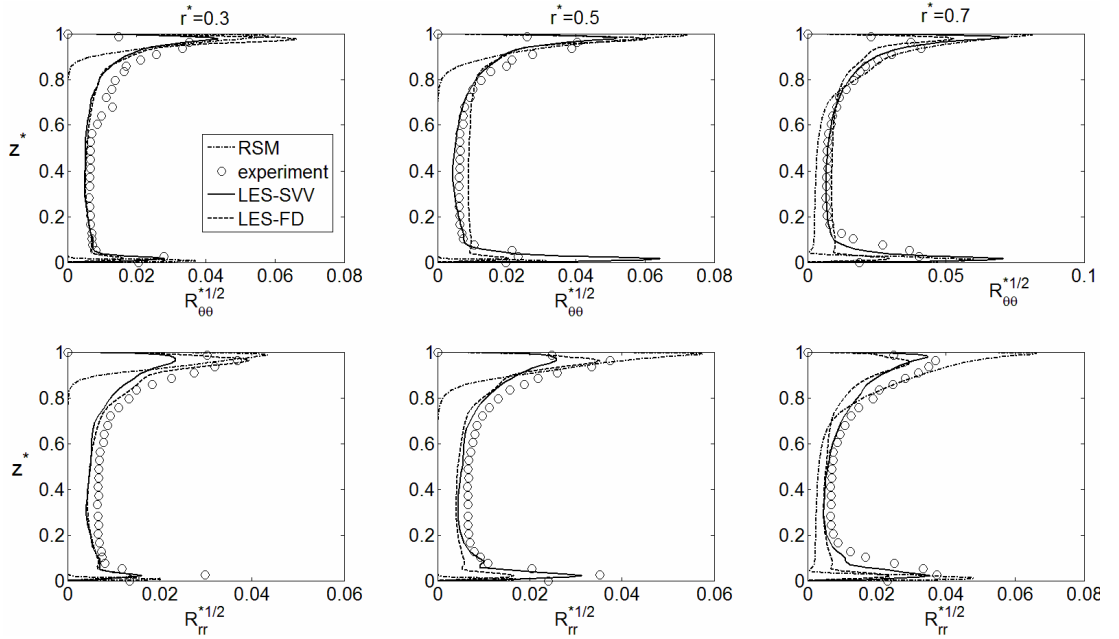

**Fig. 4** Axial profiles of the two main Reynolds stress tensor components $R_{rr}^*$ and $R_{\theta\theta}^*$ at three radial locations. Comparisons between the LES-SVV (full lines), the LES-FD (dashed lines), the velocity measurements (circles) and the RSM model (dash-dotted lines).

Second-order statistics available from experimental measurements in the radial $R_{rr}^* = \overline{v_r'^2}/(\Omega r)^2$ and tangential directions $R_{\theta\theta}^* = \overline{v_\theta'^2}/(\Omega r)^2$ have been computed in Figure 4 at the same radial locations.

LES results provide an overall agreement with the experimental data both in boundary layers and in the core with a slightly better estimation of the turbulence intensity by LES-SVV than LES-FD. Surprisingly, RSM seems to strongly overestimate the maxima of the normal components of the Reynolds stress tensor within the stator boundary layer at all locations while it underpredicts the turbulence intensity in the core and completely misses the transition in the rotor layer.

The peaks of the normal Reynolds stress tensor components are relatively well predicted by both LES within the stator layer at a wall distance of $0.05h$ and $0.025h$ for the radial and tangential component, respectively. Both LES models overpredict $R_{\theta\theta}$ in both boundary layers, the maximum being reached by the LES-SVV within the rotor layer at mid-radius. That leads to a much stronger anisotropy of the Reynolds stress tensor than in experiments. Such behaviour could be related to the anisotropy of the grid computation, which is globally much coarser in the tangential direction, especially at large radii. This hypothesis is supported by the LES of Wu and Squires (2000) which observed a strong sensitivity of this component depending on the resolution.

Figure 5 illustrates the $R_{r\theta}^* = \overline{v_r'v_\theta'}/(\Omega r)^2$ shear stress at mid-radius. The other components are not illustrated because the two other cross components were not available using the LDV technique. Both LES methods show a global agreement in terms of sign and intensity within the stator boundary layer. On the rotor side, however, LES-FD underpredicts its intensity due to a lower turbulence level. As expected from the literature, the shear stress magnitude is much smaller than the normal components (see DNS by Lygren and Anderson 2001). This is a feature of rotating disk boundary layers that indicates an important structural change in the turbulence compared to the more classical plane boundary layer.

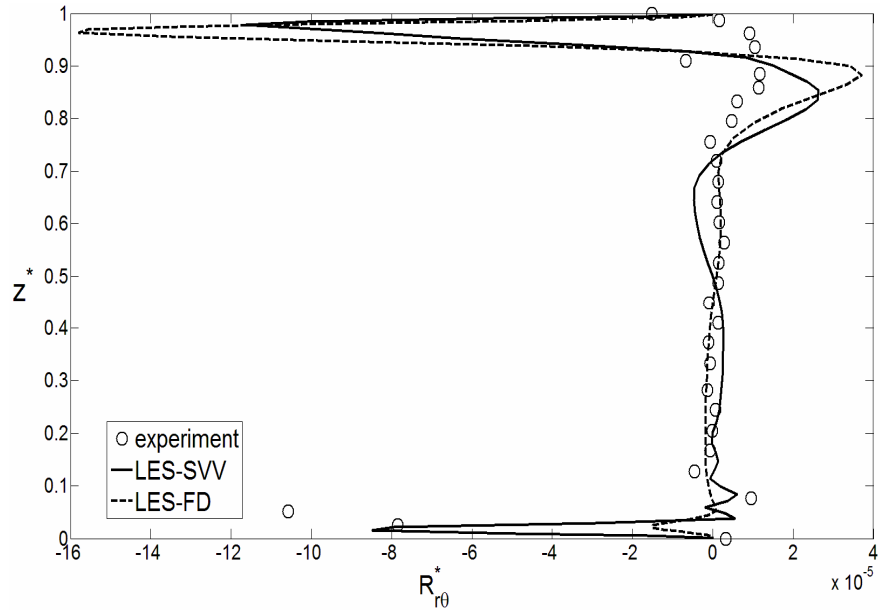

**Fig. 5** Axial profiles of the $R_{r\theta}^*$ shear stress at mid-radius. Comparisons between the LES-SVV (full lines), the LES-FD (dashed lines) and the velocity measurements (circles).

Instantaneous results also show a good agreement between both LES methods that provide almost similar coherent structures. Vortical structures are identified using positive isosurfaces of the Q-criterion.

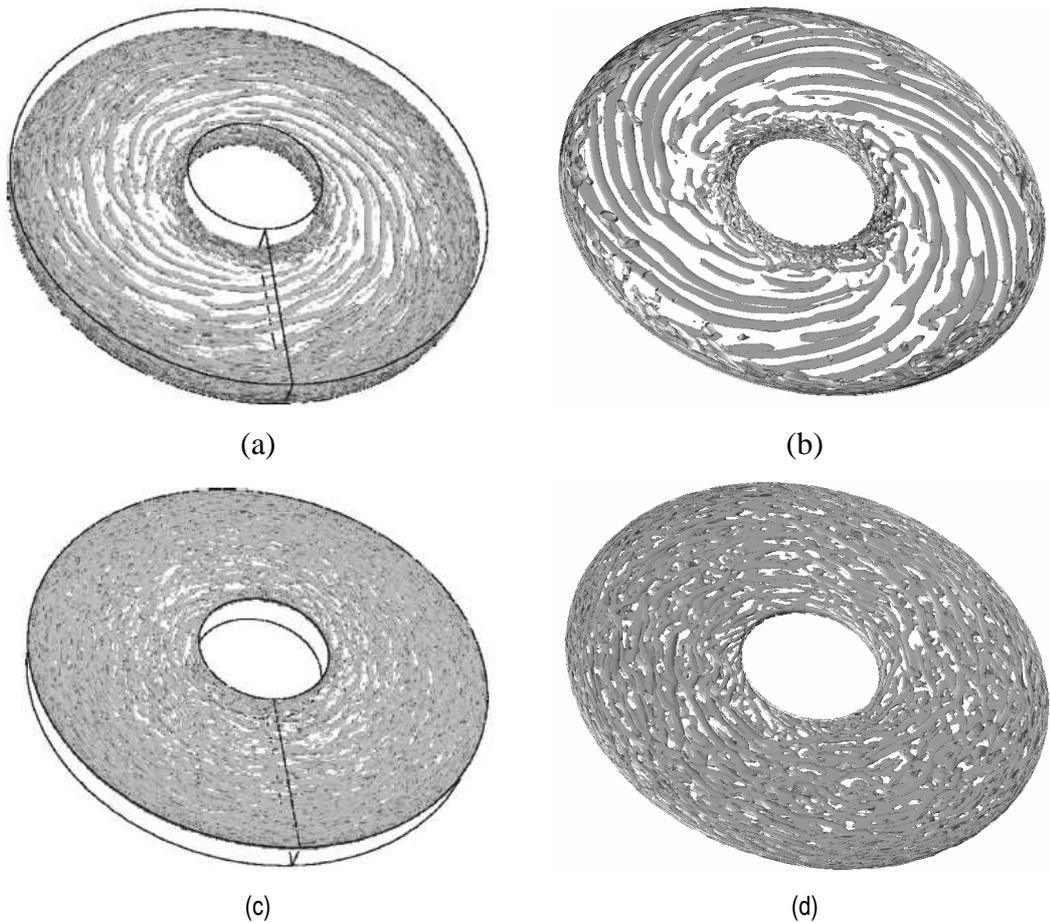

**Fig. 6** Isosurfaces of the Q-criterion along the rotor (a, b) and the stator (c, d); (a, c) LES-SVV and (b, d) LES-FD. The rotating disk rotates anticlockwise.

As already observed in the turbulence statistics, the rotating disk layer is only weakly turbulent. This is featured in the flow structure by coherent negative spiral arms (as they roll up in the opposite rotation sense of the disk) at intermediate radii (Figures 6a & b) forming an angle of about 15° with the tangential direction. This feature is characteristic of the viscous linear instability referred to in the literature as Type II and is known to play an important role in the transition process to turbulence (see Serre *et al.* 2004).

Around the hub, where the flow coming from the stator impinges the rotor, both simulations predict a highly turbulent region with smaller disorganized structures. The LES-SVV gives the transition to turbulence at the right threshold expected from theory and experiments at large radii, that is at a local Reynolds $Re_r = 386$. These spiral arms break into much smaller and more concentric structures similar to the ones observed along the stator.

Along the stator, both LES exhibit very thin coherent vortical structures aligned with the tangential direction (Figure 6). This is typical of a turbulent rotating boundary layer since the anisotropy invariant map shows that turbulence tends to the axisymmetric limit in this flow region (Séverac *et al.* 2007). The thinner structures predicted by the LES-SVV confirms nevertheless a globally higher level of turbulence.

A more accurate description of such coherent structures can be provided by isosurfaces of the Q-criterion in a ($r$, $\theta$) plane using a Cartesian frame of reference (Figure 7). At the junction between the outer cylinder and the stator (top of Figure 7), very thin structures already aligned with the tangential direction are observed. When these structures move towards the hub, convected by the mean flow, they eventually merge and get larger and very elongated due to the decrease of the turbulence intensity with the radius. Finally, in the vicinity of the hub they get inclined and stretched due to the centrifugal force induced by the rotating hub.

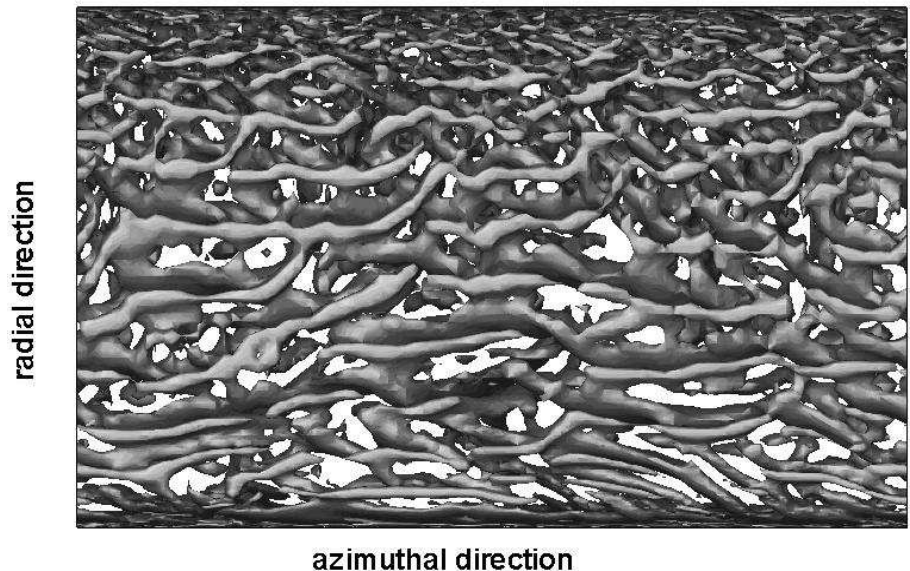

**Fig. 7** Q-criterion isosurface on the stator side view from the top using a Cartesian frame of reference (LES-FD). The rotation sense is from the left to the right (the shroud at the top, the hub at the bottom)

Due to the confinement of the cavity in the radial direction, the mass conservation involves a high streamline curvature at the corners and the occurrence of vertical boundary layers along the vertical endwalls as shown in Figure 1. Both boundary layers are centrifugally unstable and play a crucial role in the global rotor-stator flow stability. They are poorly documented, however, in the literature. The combination of the axial flow with the rotation induces two strong swirling jets that impinge the disks. The impact regions are characterized by a local increase of turbulence intensity and by much smaller and much more disorganized flow structures as clearly identified in figure 5a.

Perturbations originating from the nearly turbulent rotor layer disturb the unstable boundary layer along the outer cylinder leading to coherent structures of strong intensity, almost aligned in the θ-direction (the cylinder being fixed) as depicted on Figure 8. On the opposite side along the rotating hub, structures take the form of spiral arms with an inclination of 32° due to the influence of the rotating disk.

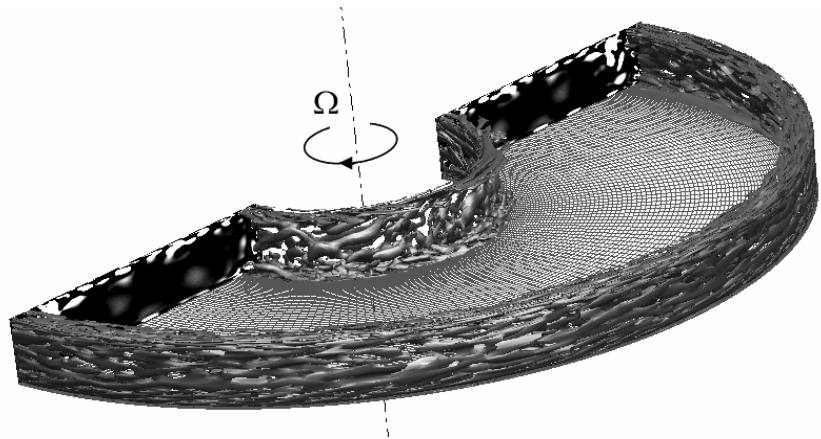

**Fig. 8** Isosurfaces of the Q-criterion along the hub and the shroud together with isolines of the Q-criterion within two ($r$, $z$) planes (LES-FD). The structures along the rotor and the stator are blanked.

Finally, Figure 8 shows that almost all the coherent structures are confined in both disk boundary layers within the meridian plane.

## 5 Concluding remarks

In this paper, we have presented a direct comparison of two LES methods for the rotor –stator flow at moderate Reynolds number Re = $4\times10^5$. These two LES methods are among the first to represent the main features of this turbulent flow. Both are based on high-order approximations, either spectral or fourth-order compact finite-difference. In LES-SVV, the energy cascade has been modelled as an increased diffusion on the highest resolved wavenumbers by the inclusion in the Navier-Stokes equations of an additional viscous operator which is homogeneous to the classical diffusive one. LES-FD is based on a Smagorinsky subgrid scale modelling, the near wall energy motions being fully resolved within a dynamic procedure.

This paper firstly shows that LES predictions may provide an overall agreement with experiments. Both simulations predict the main features of the flow with a fully turbulent stator boundary layer and a transitional rotor layer. Coherent structures under the form of spiral arms or circles within the rotor and the stator layer respectively are also well captured. LES-SVV provides a better overall agreement. In particular, the turbulence intensity is closer to that measured experimentally. Moreover, LES-SVV also predicts the location for the transition to turbulence along the rotor quite well, whereas it is missed by LES-FD. Both models, however, appear to be slightly too dissipative, leading to an underestimation of the turbulence within the cavity. This is clearly illustrated by thinner computed boundary-layers than experimentally measured ones as well as a slightly smaller tangential velocity within the core. This implies the use of thinner meshes to more accurately resolve the energetic structures of the flow even if the turbulence level in experiments is certainly artificially amplified by uncontrolled noise like vibrations or roughness. This requires the improvement of the currently used algorithms in terms of cost reduction through a massive parallelization of the codes.

In conclusion, the rotor-stator case characterized by a strong inhomogeneity of the flow and thin boundary layers with persisting large-scale three-dimensional precessing vortices strains all the components of a LES solver as confirmed by the very limited literature on the topic. Nevertheless, the comparisons between the present results and those obtained by RANS using a modelled designed for rotation show that LES is the right level of modelling for such flows. Finally, the overall agreement between experiments and our two LES models using subgrid scale models free of any ingredients for rotation offers indirect support to the idea that when using high-order numerical methods (low numerical dissipation) the influence of the LES modelling in complex flows seems weaker. Such behaviour should encourage the LES community to increase its effort on SGS modelling and in the use of high-order schemes.

**Acknowledgements**
LES-SVV was granted access to the HPC resources of IDRIS under the allocation 2009-0242 made by GENCI (Grand Equipement National de Calcul Intensif). The LES-FD calculations have been performed on the M2P2 cluster composed of 2 Xeon quadcore 3 GHz. The work was supported by CNRS GDRE-MFN in the frame of the DFG-CNRS program "LES of complex flows".